\documentclass[12pt]{article}

\usepackage{sbc-template}
\usepackage{graphicx,url}
\usepackage[utf8]{inputenc}

\usepackage[tableposition=top]{caption}
\usepackage{svg}

\usepackage{array}
\usepackage{multirow}
\usepackage{hhline}
\usepackage{mhchem}

\usepackage{hyperref}

\newcommand{\arquitetura}{\textit{pmSensing}}

\title{\arquitetura: A Participatory Sensing Network for Predictive Monitoring of Particulate Matter \thanks{This work was carried out with financial support from the FAPES, FAPEMIG, FAPESP (Grants \#2018/23011-1 e \#2020/05182-3),  CNPq and Coordenação de Aperfeiçoamento de Pessoal de Nivel Superior – Brasil(CAPES) - Código  001.}}

\author{Lucas L. S. Sachetti\inst{1},  Enzo B. Cussuol\inst{1}, José Marcos S. Nogueira\inst{2}, Vinicius F. S. Mota\inst{1}}

\address{Departamento de Informática  -- Universidade Federal do Espírito Santo\\
   Vitória  -- Brasil
  \vspace{-0.3cm}
  \email{ \{lucas.sachetti,enzo.cussuol\}@edu.ufes.br  vinicius.mota@inf.ufes.br}
  \nextinstitute
  Departamento de Ciência da Computação -- Universidade Federal de Minas Gerais\\
  Belo Horizonte  -- Brasil
  \vspace{-0.3cm}
  \email{jmarcos@dcc.ufmg.br}
}

\begin{document} 

\maketitle

\begin{abstract}

This work presents a proposal for a wireless sensor network for participatory sensing, with IoT sensing devices developed especially for monitoring and predicting air quality, as alternatives of high cost meteorological stations.  The system, called \arquitetura, aims to measure particulate material. A validation is done by comparing the data collected by the prototype with data from stations. The comparison shows that the results are close, which can enable low-cost solutions to the problem. The system still presents a predictive analysis using recurrent neural networks, in this case the LSTM-RNN, where the predictions presented high accuracy in relation to the real data.
\end{abstract}

\section{Introduction}

Air pollution has been a constant concern for the population. Because of this, the monitoring of data on air quality has gradually increased.
The main pollutants monitored are Carbon Monoxide (CO), Ozone (O$_{3}$), Nitrogen Dioxide (NO$_{2}$), Sulfur Dioxide (SO$_{2}$) and o Particulate Material(PM) ~\cite{roy2017smart, dos2018emprego}. 
Monitoring of these pollutants is usually done on a macroscale by meteorological stations at strategic points. However, weather stations have a high cost of installation and maintenance ~\cite{iema2021}.

Recent works propose participatory sensing and/or Internet of things for monitoring polluting gases at a low cost ~\cite{liu2015bicycle, dos2018emprego}. However, the collection of particulate matter is usually left to expensive sensors, such as those available at meteorological stations. The particulate material varies in quantity, size, shape and chemical composition, being measured as particles with a diameter of: up to 10 $\mu$m (PM$_{10}$), less than 2.5 $\mu$m (PM$_{2.5}$) and less than 1 $\mu$m (PM$_{1.0}$)~\cite{abbey1999long}. The smaller the particles, the deeper they enter the respiratory system, causing damage to health, which will depend on the composition of the particulate material~\cite{who:hobook09}.However, particulate material sensors capture any type of particle from the air, even non-toxic ones like pollens, for example. For this reason, this work considers that the increase of particulate matter sensors in a region can minimize the problem caused by the wrong sensing of only one sensor. For this, it is necessary that the particulate material sensor network has a low cost but high reliability.


{This work presents a proposal for a wireless sensor network for participatory sensing, with IoT sensing devices specially developed for monitoring and predicting air quality.
In the case of this paper, the sensor device, called \arquitetura, specifically aims to collect particulate material for monitoring and predicting its occurrence in the monitored environment. To this end, a prototype of the device was developed using low-cost sensors, processing and storage capacity, isolated operation or connected to the Internet with a Wi-Fi interface for accessing cloud platforms.
Due to the low cost, the number of sensors around the city can be increased, even allowing citizens to install them in their homes, thus creating a network of participatory sensing and monitoring. In this way, the \arquitetura~aims to enable a broader and even mobile or dynamic coverage of points in a city, allowing a micro-scale view of air quality, towards the creation of smart cities. For example, the \arquitetura~ could be attached to the city's shared buses and/or bicycles for continuous and participatory monitoring of air quality. In addition to the sensing network, the proposed architecture aims at actively monitoring the air quality, that is, verifying that the amount of particulate matter at a given time is within the expected range. For example, authorities could reduce vehicle traffic at times predicted for high pollution in a region.
}


Initially, a validation of the prototype was carried out, comparing the monitoring carried out by the \arquitetura~ network with the monitoring carried out by sensors in meteorological stations. After that, the prototype of the device was made available for non-specialist users to monitor their home environment. The monitored data was periodically sent to a cloud platform. Then, the accuracy of prediction models from the data collected by \arquitetura was analyzed.
The main contributions of this work are:
\begin{itemize}
    \item Proposition and design of a sensor network with participatory monitoring for environmental monitoring of particulate matter;
    \item Development of a sensing prototype with wireless collection, storage, pre-processing and communication capabilities;
    \item Demonstration that the use of low-cost sensors used by volunteer users can reach results similar to those obtained by high-cost sensors from meteorological stations.
\end{itemize}

The remainder of this work is organized as follows: Section \ref{sec:funteo} presents a theoretical foundation along with related work.
Section \ref{desenvolvimento} presents the development, operation, validation and implementation of the \arquitetura.
  Section \ref{results-analise} presents an evaluation of the system. Section \ref{analise-preditiva} shows a predictive analysis using LSTM. Section \ref{conclusao} concludes the study by discussing results, limitations and future work.

\section{Theoretical Background and Related Work} \label{sec:funteo}
This section presents a theoretical basis on work-related themes, as well as related works that were studied during development.

\subsection{Air Quality}
The growing urbanization across the globe has generated an increase in energy consumption and in the emission of pollutants, whether from the burning of fossil fuels or from industrial production.

The main monitored pollutants that affect people's health are nitrogen oxide (NO2 or NOx), carbon monoxide (CO), sulfur dioxide (SO2), ozone (O3) and particulate matter.
Interest in the health-related effects of air pollution became more pronounced after two studies by a US group suggested that exposure to air pollution was associated with shortening people's lives \cite{dockery1993association}. In particular, significant relationships have been demonstrated between particulate matter with a diameter less than 10$\mu{m}$ (PM$_{10}$) and respiratory diseases \cite{pope1991respiratory,roemer1993effect} and even with deaths from non-respiratory diseases malignancies in men and women, as well as with lung cancer mortality in non-smoking men \cite{abbey1999long}.

There are public policies for the monitoring of gases and particulate materials using meteorological stations. In the state of Espírito Santo, the State Institute for the Environment and Water Resources (IEMA) provides an air quality index according to the concentrations of pollutants, shown in the Table \ref{tabclass} at $\mu$g/$m ^3$ (microgram per cubic meter). The relationship between the amount of pollutants measured by the sensors and the effects on health are shown in Table \ref{tabrisco}. It is noteworthy that the database collected by IEMA is public and the direction of this work in monitoring PM$_{2.5}$ is also due to the availability of these data.

\begin{table}[htb]
    \centering
    \scalebox{0.8}{
    \begin{tabular}{|c|c|c|c|c|c|c|} \hline
        \multirow{3}{*}{Classification} & \multicolumn{6}{c|}{Concentration} \\\hhline{~------}
            & \ce{PM10} & \ce{PM_{2.5}} & \ce{SO2} & \ce{NO2} & \ce{O3} & \ce{CO} \\
            & 24h & 24h & 24h & 1h & 8h & 8h \\\hline
        Good & 0 - 50 & 0 - 25 & 0 - 20 & 0 - 200 & 0 - 100 & 0 - 10.000 \\\hline
        Moderate & $>$50 - 120 & $>$25 - 60 & $>$20 - 60 & $>$200 - 240 & $>$100 - 140 & $>$10.000 - 13.000 \\\hline
        Unhealthy & $>$120 - 150 & $>$60 - 125 & $>$60 - 365 & $>$240 - 320 & $>$140 - 160 & $>$13.000 - 15.000 \\\hline
        Very Unhealthy & $>$150 - 250 & $>$125 - 210 & $>$365 - 800 & $>$320 - 1130 & $>$160 - 200 & $>$15.000 - 17.000 \\\hline
        Hazardous & $>$250 & $>$210 & $>$800 & $>$1.130 & $>$200 & $>$17000 \\\hline
    \end{tabular}}
    \caption{Air quality by pollutant concentration\cite{iema2021}.}
    \label{tabclass}
\end{table}

\begin{table}[htb]
    \centering
     \scalebox{0.8}{
    \begin{tabular}{|c|!\raggedleft m{11cm}|} \hline
        Quality & \multicolumn{1}{c|}{Meaning} \\\hline
        \begin{tabular}{c}Good\\ or\\ Moderate \end{tabular} & People from sensitive groups (children, the elderly and people with cardiorespiratory diseases) may present symptoms such as dry cough and tiredness. The general population is not affected. \\\hline
        Unhealthy & The entire population may present symptoms such as dry cough, tiredness, burning eyes, nose and throat. People from sensitive groups can have more serious health effects. \\\hline
        Very Unhealthy & The entire population may present worsening of symptoms such as dry cough, tiredness, burning eyes, nose and throat, as well as shortness of breath and wheezing. Even more serious effects on the health of sensitive groups. \\\hline
        Hazardous & The entire population may be at risk of manifestations of respiratory and cardiovascular diseases. Increase in premature deaths in people from sensitive groups. \\\hline
    \end{tabular}}
    \caption{Relationship between air quality and population health risks}
    \label{tabrisco}
\end{table}

\subsection{Time Series Predictions}
\label{subsec-lstm}

The data collected by the sensors form a set of observations ordered in time, called time series~\cite{azzouni2017long}. The data collected by the sensing proposed in this work are treated as a time series, which allows the use of analytical methods to calculate the air quality, whether good or not. These methods include: descriptive analysis, which only describes current and past events; predictive analytics, which uses past data to predict future actions; prescriptive analysis, to make or recommend decisions based on the results obtained by predictive and/or descriptive analysis. In the context of this work, predictive analytics enables active monitoring of air quality, identifying observations outside the expected values, enabling, for example, the generation of alerts for competent authorities.

There are several ways to predict data, including the use of Artificial Neural Networks (ANN) \cite{crone2010feature}, commonly used for pattern recognition. A Recurrent Neural Network (RNN) uses data that has a strong relationship with time \cite{bone2003boosting}. RNNs use past information to support current data in the neural network. When data are strongly related to the recent past, the Long Short-term Memory Recurrent Neural Network (LSTM) are more effective in making predictions \cite{azzouni2017long}. An LSTM not only retains the timing characteristics of the RNN structure, it also has the memory function for time series.

The original LSTM idea was initially proposed in \cite{hochreiter1997long} and since then different models have been proposed to improve its performance \cite{cho2014learning}, \cite{li2019ea}, \cite{karevan2020transductive}, \cite{hu2020time}. LSTMs use a locking mechanism to define the data, its duration and read time in the memory cell.

The LSTM-RNN consists of a module with memory cells that can learn time-domain data characteristics. The memory module in the LSTM recurrent neural network contains three gates: an input gate, a forget gate, and an output gate. These ports control the input, updating and output of information, respectively, so that the network has a certain memory function. The output is a number between 0 and 1, where 1 indicates keeping all input information, and 0 indicates removing that information. This work proposes the use of LSTMs to predict particulate matter from time series collected by sensors of the proposed system.

\subsection{Related Works}

Proposals for applications, protocols and performance analysis related to the Internet of Things applied to air quality monitoring, mainly in smart cities, are widely discussed in the literature.~\cite{da2015micro,dos2018emprego}. 

\cite{da2015micro} developed a low-cost sensor to monitor air quality by sending information via Bluetooth communication to a cell phone running the Android system, to measure concentrations of NH3, NOx, benzene, smoke and CO2. At work \cite{liu2015bicycle} public bicycles were used to collect data on pollution in the city, measuring gases and particles, using Bluetooth technology (FBT06M) to send collected data to a base station.

In \cite{dos2018emprego} a kit of sensors installed in buses in a city was used to identify the regions with the highest levels of pollution. The gases $O_3$, $NO_2$, $CO_2$, the ambient temperature and the humidity of the air were measured. A city map was constructed showing the intensity of pollution at each point on the bus route. In \cite{al2010mobile}, the authors created a system with a device with sensors to measure the air quality (CO, NO2 and SO2) of a city, send the data via the {mobile} network to a server and build a map of the pollution. The kit uses a GPRS modem to send data to a server that uses a Google Maps API. The initial proposals of the \arquitetura~ were precisely the installation of the prototype on shared buses and bicycles, but due to bureaucratic issues and the short term, they were discontinued.

Regarding the prediction of PM2.5 concentration using LSTM networks, which is a technique also used in this work, the works of \cite{thaweephol2019long} and \cite{li2020urban} stand out. In \cite{thaweephol2019long}, the authors trained an LSTM network model using time series data on air quality from a metropolitan police station in the city of Bangkok. Their results were compared with another model used called SARIMAX (Seasonal AutoRegressive Integrated Moving Average with eXogenous regressor), which is an extended version of the traditional ARIMA model. The comparison, made using error metrics, shows that the SARIMAX model obtained twice as many errors in relation to the LSTM. The authors showed a strong correlation between PM2.5 and other pollutants such as PM10, NO and CO.

In \cite{li2020urban} developed the Attention-based CNN-LSTM, a variant of the LSTM with convolutional networks (CNN-LSTM) that uses a layer with attention mechanisms to capture the most important parts of the concentration of PM$ _{2.5}$ when considering characteristics of past states. Some correlations between particulate matter concentrations and air temperature, humidity and visibility (km) are also shown. The authors compared their proposal with the Support Vector Regression (SVR), Multilayer Perceptron (MLP), Random Forest Regression (RFR), RNN, LSTM, and CNN-LSTM, and showed which obtained satisfactory results in relation to the Root Mean Square Error (RMSE) and Mean Absolute Error (MAE) error metrics.

\section{Development, operation, validation and implementation of \arquitetura} \label{desenvolvimento}

This section presents the sensing \arquitetura module, its architecture, its functioning, the way in which the raw data collected are transformed, as well as the implantation of units in a region of the city.

\subsection{Architecture and components}

\begin{figure}
\centering
\includegraphics[width=0.7\textwidth]{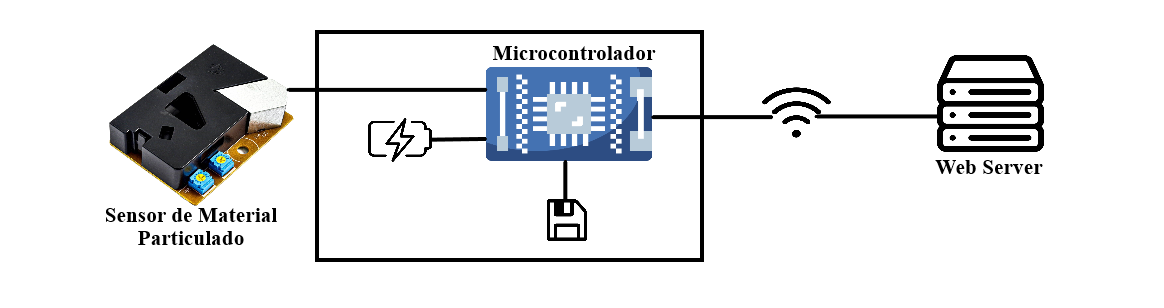} 
\caption{Sensing module architecture}
\label{arquitetura2}
\end{figure}

The sensing module \arquitetura, whose structure is shown in Figure \ref{arquitetura2}, is composed of a microcontroller with memory, a particulate material sensor and a wireless communication interface. For the development of the module, cost, energy consumption, memory and performance were taken into account. For this version, the NodeMCU ESP8266 microcontroller and 802.11 wireless network technology were used.

\begin{figure}[!ht]
\centering
\begin{minipage}{.35\textwidth}
  \centering
  \includegraphics[width=0.9\linewidth]{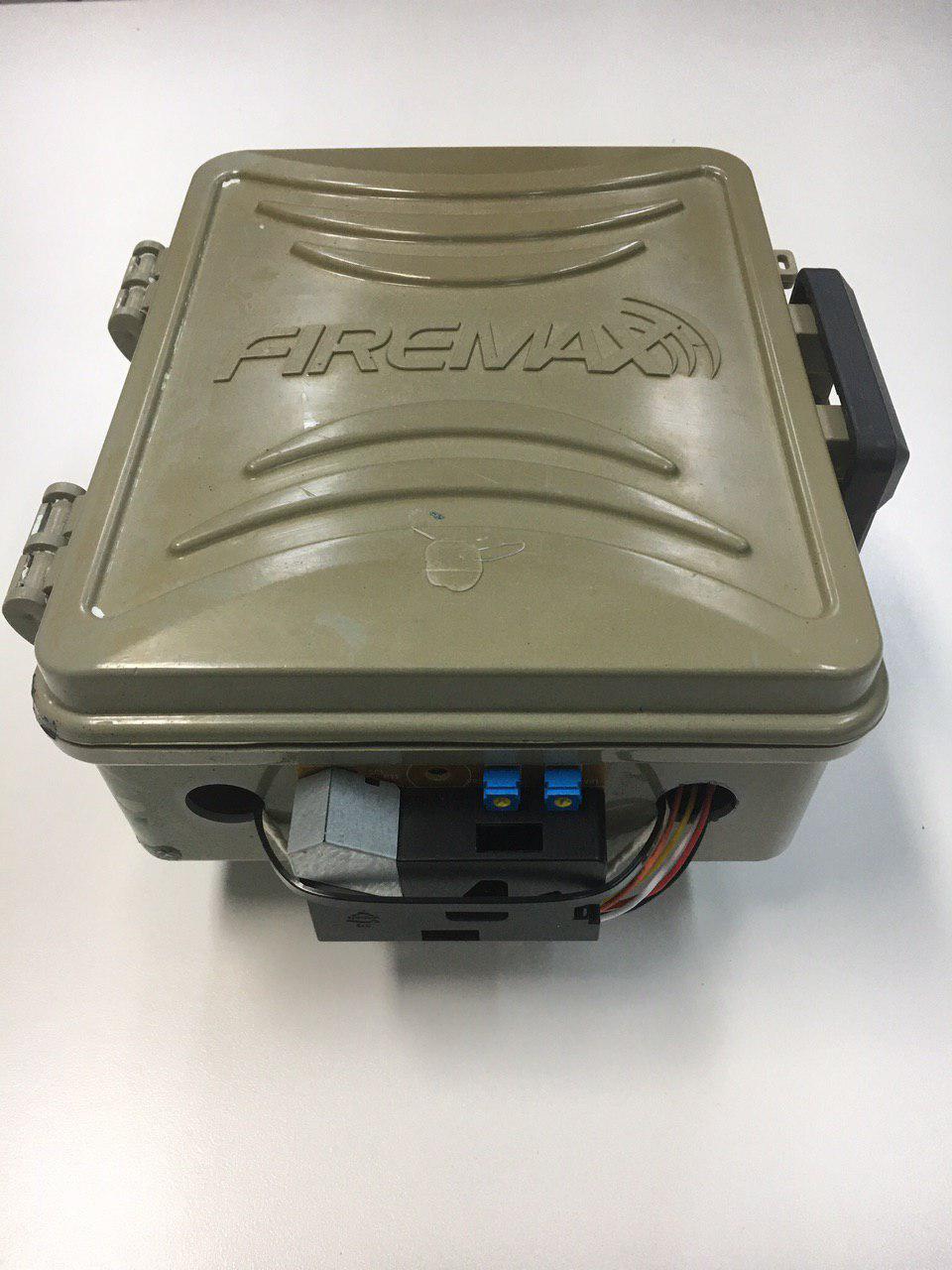}
  \captionof{figure}{Closed sensor prototype}
  \label{fechado}
\end{minipage}%
\begin{minipage}{.35\textwidth}
  \centering
  \includegraphics[width=0.9\linewidth]{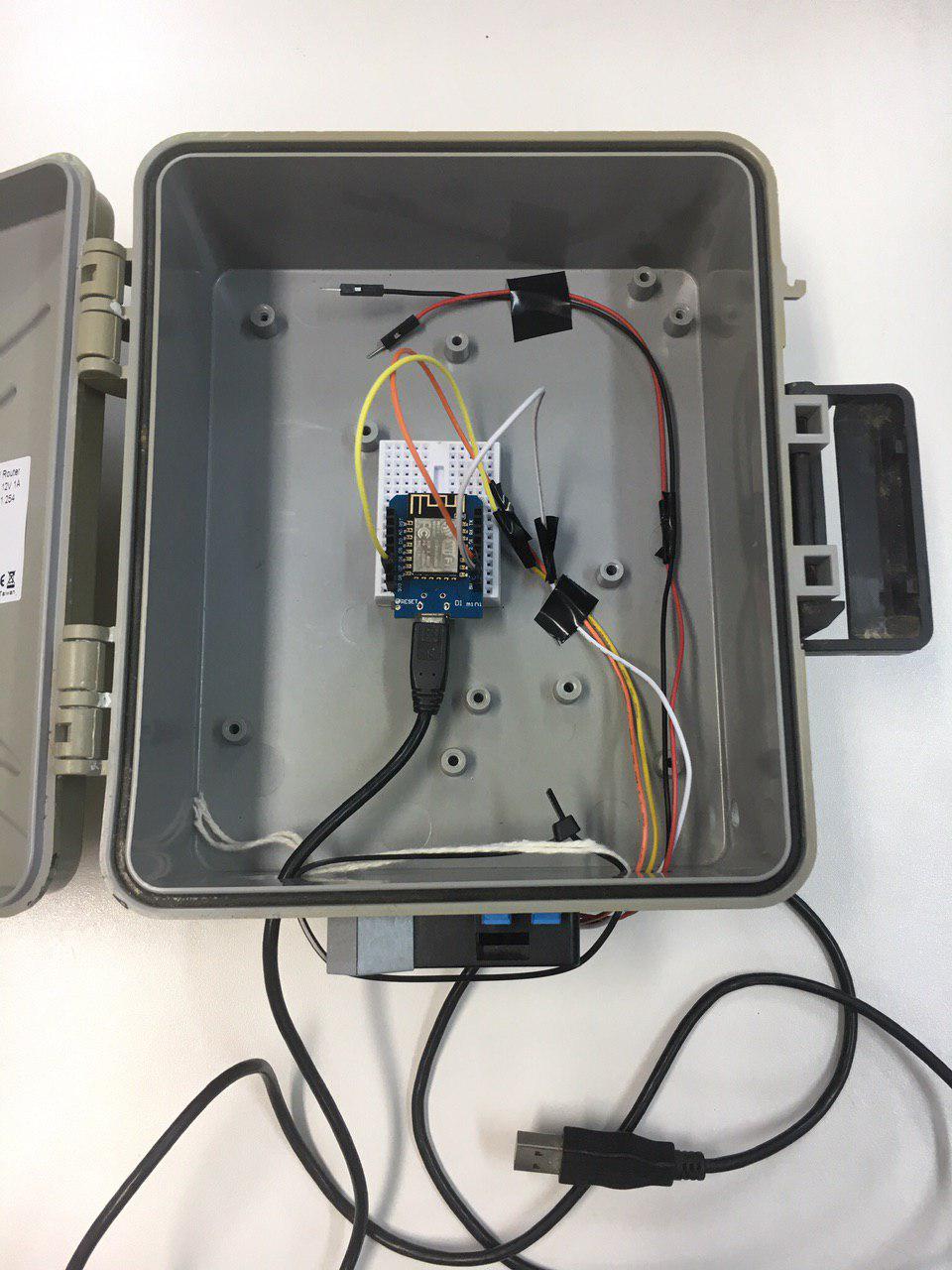}
  \captionof{figure}{Open sensor prototype}
  \label{aberto}
\end{minipage}
\end{figure}

As the stations provided by the Federal University of Espírito Santo (UFES) and by IEMA are located in an open environment, subject to rain, wind and sun, among other factors, the prototype was stored in an airtight box. The microcontroller and battery were stored internally, and the particulate matter sensor was installed on the outside of the box. Figures \ref{fechado} and \ref{aberto} show the prototype that was developed.

To operate in external environments without the possibility of accessing the internet to send collected data, the solution uses a microcontroller with non-volatile memory. For this version of the \arquitetura~ a low energy cost microcontroller was chosen, with local storage capacity and an available 802.11 network. Connectivity can be expanded using other communication technologies such as GSM or even LoraWan.
On the other hand, the local storage needs to be enough to collect data during the offline period. The frequency of data reading and storage is a configurable parameter of the proposed system.  

\subsection{Operation}

The system operates offline, for monitoring environments without internet connection, and online, for environments protected from the weather and that have wireless networks. Offline mode is also used to validate sensors with stations that are located in disconnected environments. Validation allows verifying whether the monitoring of the \arquitetura~ approaches the results of the stations.

In online mode, for operation in protected environments, data is sent via the network to a web server, using wireless access that exists on site. Data are sent to an open platform which, in this work, is the ThingSpeak \cite{thingspeak} platform for data storage and analysis.

With the sensing module in online mode, the data collected and pre-processed in the module itself are sent to the ThingSpeak platform to be stored and processed. This platform allows the creation of channels that receive data via web socket or via the MQTT protocol. Every 30 seconds data from a particulate matter sensor reading is sent to the platform. This time is the minimum interval for sending messages defined by the platform in the free plan.

\subsection{Collection and pre-processing of data}

The particulate material sensor used was the Samyoung DSM501 dust sensor, which is a low cost and compact size particle density sensor manufactured by the Korean company Samyoung Electronics. The sensor has a heater (resistor) to generate heat that creates an upward current of air that draws outside air into the module. The DSM501 measures the length of time a particle is detected by the sensor. This time is called the Low Pulse Occupancy, which can also be seen as the percentage of opacity of the air flowing through the sensor. In this way, it detects dust, particles and pollen in the air ( $PM_{2.5}$ and $PM_{1.0}$) indoors.
The \arquitetura~ monitors $PM_{2.5}$, making it possible to validate and compare the data obtained with data from IEMA stations.

In order for the data comparison to take place, it is necessary to convert the sensed data to another format, concentration in micrograms per cubic meter, in this case called Low Ratio. According to the manufacturer's datasheet, the raw data obtained by the sensor is in the form of Low Pulse Occupation. The conversion is done in the module itself. Equation \ref{eq2}, converts the Low Pulse Occupation to a concentration of particles per liter (pcs/L) or particles per 0.01 cubic feet (pcs/0.01
cf) (Low ratio [\%]).

\begin{equation}
\label{eq2}
Concentration = 1.1\times( \textit{Low ratio}^3) - 3.8 \times(\textit{Low ratio}^ 2) + 520\times\textit{Low ratio} + 0.62
\end{equation}
This concentration can be converted into concentration per cubic meter ($\mu g/m^3$), applying the Equation \ref{eqref}, as shown in \cite{roy2017smart}.

\begin{equation}
\label{eqref}
\begin{split}
    Particle_{Mass} = & Density\times Volume \\
    PM_{Concentration}(\mu g/m^3) = & Number\_Of\_Particle \times 3531.5 \times Particle_{Mass} \\
\end{split}
\end{equation}

 Importantly, \cite{roy2017smart} assume that all particles of this concentration are spherical and have a density equal to 1.65$\times10^{12}$ and that the radius of each particle, in the case $PM_2.5$, is equal to 0.44 $\mu$m. The mass is made up by multiplying the density by the volume, and the concentration is obtained from the multiplication of this mass by the number of particles and by a measurement conversion factor.

\subsection{Implementation of sensing units} \label{implantacao}

The sensors were deployed in locations close to the stations of the State Institute for the Environment and Water Resources (IEMA) of Espírito Santo in Grande Vitória-ES. In this way, the collected data were compared with the data collected by the IEMA macroscale sensors. The data provided by IEMA are from collections made every 60 minutes; therefore, to allow for an adjusted validation, the mean concentration of particulate matter for every 60 minutes was calculated.

\begin{figure}[!ht]
\centering
  \includegraphics[width=0.35\linewidth]{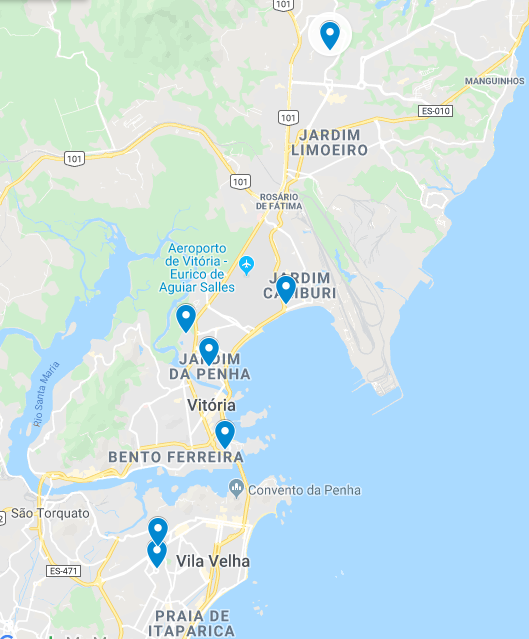}
  \captionof{figure}{Map of monitoring sites}
  \label{mapa1}
\end{figure}

To assess participatory sensing, four prototype units were developed and distributed to volunteers. To collect data in more places, prototypes were handed out to new volunteers. Figure \ref{mapa1} shows the locations where the sensors were installed for data collection. The more people using it in a region, the better the median of the results that will be more reliable to the results of the meteorological stations in which they are used for validation.


\section{\arquitetura~evaluation}\label{results-analise}

The device prototype, built with low-cost particulate material sensors, was evaluated by comparing particulate material measurements taken by meteorological stations. The data collected by the sensing units at various points in Grande Vitória, with the help of volunteers, were evaluated and classified according to the Air Quality Index (AQI), made available by IEMA.
The sensors were placed in houses close to IEMA monitoring stations.

\subsection{Validation}\label{validacao}

In preliminary tests, it was observed that the positioning of the prototype is relevant to the results. However, as long as it is placed in positions exposed to the weather, such as near a window or on a balcony, the results are satisfactory in relation to the data observed by the meteorological stations, as shown below.

\begin{figure}
\centering
\begin{minipage}{.45\textwidth}
  \centering
  \includegraphics[width=1\linewidth]{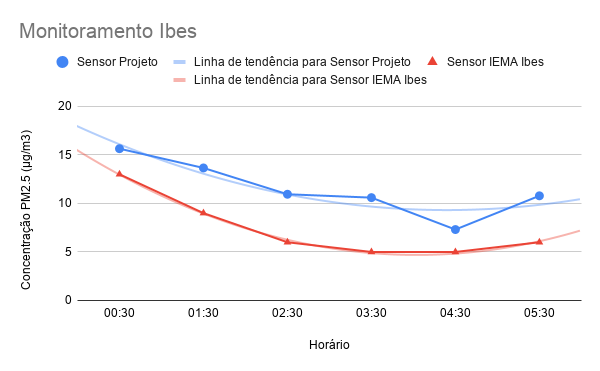}
  \captionof{figure}{Monitoring of $PM_2.5$ in the Ibes neighborhood}
  \label{ibes}
\end{minipage}%
\begin{minipage}{.45\textwidth}
  \centering
  \includegraphics[width=1\linewidth]{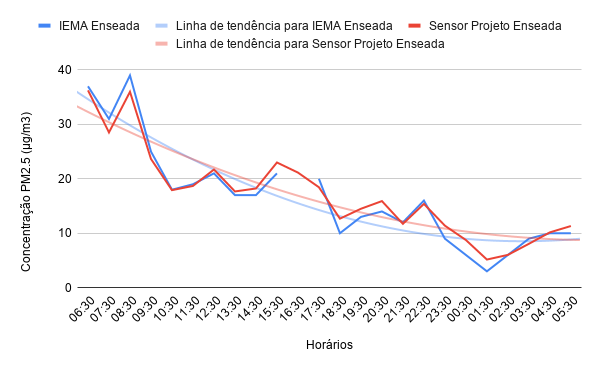}
  \captionof{figure}{Monitoring of $PM_2.5$ in Enseada do Sua}
  \label{enseada-validacao}
\end{minipage}
\end{figure}

 The first validation test took place in May 2019 (05/28/2019) in a location close to the IEMA macroscale sensor in the Ibes neighborhood in Vila Velha. Figure \ref{ibes} presents the concentration data of $PM_{2.5}$~in $\mu g/m^3$ collected by the \arquitetura~ in comparison with the macroscale sensor data from IEMA. It is observed that the \arquitetura~obtained a concentration of 3.22 $\mu g/m^3$ on average compared to the concentration measured by the IEMA. Although there is this difference in reading, both monitored data are classified as GOOD in the Air Quality Index, according to the Table \ref{tabclass}. Furthermore, it is observed that the trend curve has a similar behavior.

The second validation test was carried out near the IEMA station at Enseada do Suá, in Vitória. This lasted 24 hours and took place on August 28 and 29, 2019. The \arquitetura~ was installed on a balcony of an apartment on the sixth floor of a building next to the Fire Department where the IEMA sensor was installed. Figure \ref{enseada-validacao} shows the result of the monitoring carried out by the \arquitetura~ of the IEMA monitoring for that day. It is observed that the data coincide in many of the points observed and the average difference between the series was 0.53. The database for the day of collection was incomplete, and therefore there was a discontinuity of the IEMA data curve at this point (4:30 pm). One of the challenges of comparing the IEMA baseline is that data for a given month is released only for the following month. Therefore, \arquitetura~performed data collection on specific days but can only compare data in the month following collection.

\begin{figure}
\centering
\begin{minipage}{.5\textwidth}
  \centering
  \includegraphics[width=0.8\linewidth]{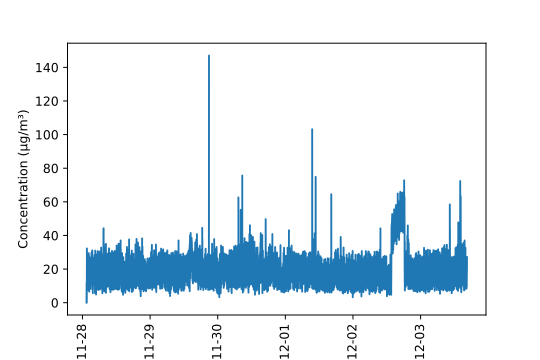}
  \captionof{figure}{PM Concentration (Jardim Camburi)}
  \label{feedteste}
\end{minipage}%
\begin{minipage}{.5\textwidth}
  \centering
  \includegraphics[width=0.8\linewidth]{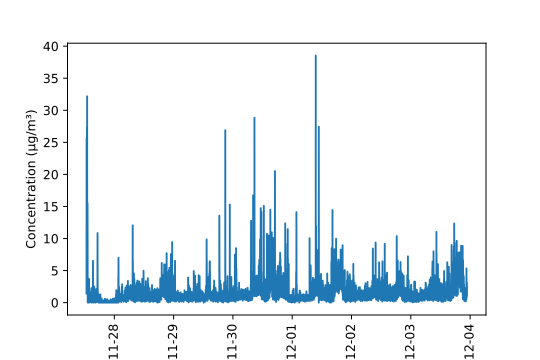}
  \captionof{figure}{PM Concentration (Jardim da Penha)}
  \label{feedcasa}
\end{minipage}
\end{figure}

\subsection{\arquitetura~as a participatory sensing system}

After validating the effectiveness of the \arquitetura~comparing it with IEMA macroscale sensors, the prototypes were used to monitor the air quality index in domestic environments. This experiment conducted in the Grande Vitória region was widely accepted by volunteers due to a popular questioning about the incidence of dust and ore dust near the port area of ore flow in the region. In neighborhoods close to this port area, IEMA monitors only PM$_{10}$ and other gases. However, the prototype of \arquitetura~monitors only PM$_{2.5}$ and PM$_{1.0}$. Therefore, there was no comparison with data from the IEMA, but with data collected concurrently in different neighborhoods.

A comparison to be highlighted is between Jardim Camburi and Jardim da Penha. Figure \ref{feedteste} shows particulate matter levels between November 28th and December 3rd. The daily average in this period was 28,745 ${\mu}g/m^3$, which is considered MODERATE in the AQI in the Jardim Camburi neighborhood. You can also see that there was a peak of a few hours above 40 ${\mu}g/m^3$, which is still considered MODERATE and few peaks above 60 ${\mu}g/m^3$, considered UNHEALTHY in the AQI. It is worth mentioning that Jardim Camburi is the closest neighborhood to the iron ore export port. On the other hand, in Jardim da Penha, $\sim$7km away from the port area, measurements were taken at the same time as in Jardim Camburi. Figure \ref{feedcasa} shows the monitoring in the same period of time as the previous one, when the daily average of concentration was 3,920 ${\mu}g/m^3$, it is also possible to notice some peaks above 25 ${\mu}g/m^3$. Despite this, the AQI at this location was considered GOOD.

\begin{figure}
\centering
\begin{minipage}{.5\textwidth}
  \centering
  \includegraphics[width=0.8\linewidth]{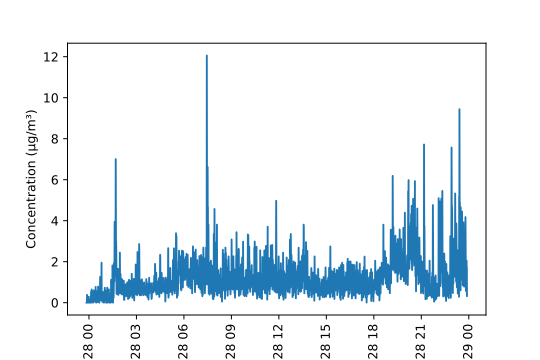}
  \captionof{figure}{PM Concentration day A (Jardim da Penha)}
  \label{feedcasaA}
\end{minipage}%
\begin{minipage}{.5\textwidth}
  \centering
  \includegraphics[width=0.8\linewidth]{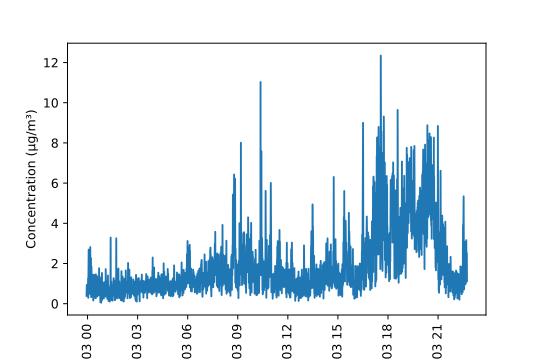}
  \captionof{figure}{PM Concentration day B (Jardim da Penha)}
  \label{feedcasaB}
\end{minipage}
\end{figure}

In order to verify the seasonality of the data, a few days of monitoring in the same place were highlighted. Figure \ref{feedcasaA} shows the monitoring in Jardim da Penha on November 28, and it is possible to see peaks around 9:00 and an increasing number of peaks after 18:00. This same increase in peaks is also observed on other days, such as the one in Figure \ref{feedcasaB} which refers to December 3rd in the same place.

Table \ref{table:2} presents the air quality index monitored at each location on the date of the experiment. It is noteworthy that the monitoring was carried out in domestic environments and should not be generalized to the entire region, a priori.

 \begin{table}[h!]
     \begin{center}
     \scalebox{0.7}{
     \begin{tabular}{|p{4cm}|p{4cm}|p{4cm}|  }
         \hline
         \textbf{Local} &\textbf{PM2.5 (${\mu}g/m^3$)} & \textbf{AQI}\\
         \hline
         Aribiri & 3.224 & GOOD\\
         \hline
         Colina de Laranjeiras & 1.042 & GOOD\\
         \hline
         Enseada do Suá & 17.178 & GOOD \\
         \hline
         Ibes & 11.369 & GOOD\\
         \hline
         Jardim Camburi & 28.745 & MODERATE\\
         \hline
         Jardim da Penha & 3.920 & GOOD\\
         \hline
         UFES & 1.515 & GOOD \\
         \hline
    \end{tabular}}
    \caption{AQI monitored in multiple locations}
    \label{table:2}
    \end{center}
    \end{table}

\section{Predictive Analytics} \label{analise-preditiva}

In order to carry out predictive monitoring, it was analyzed whether predictive models fit the particulate matter data. Predictive monitoring aims to predict whether the amount of particulate matter at a given time is within expectations, and if anomalies occur, the system could issue alerts, for example. Such monitoring can help in making environmental policy decisions.
For example, knowing in advance the levels of pollution in a location can help decision making on the flow of vehicles in Smart Cities.
This section presents a prediction model that uses data collected by \arquitetura.

\subsection{Problem definition}
The prediction of PM$_{2.5}$ is a linear regression problem and can be solved with supervised machine learning algorithms. To predict the concentration of pollutants for the next hour, $t+1$, not only the current value of the concentration, $t$, but also its previous values, $t-1$, $t-$, .., $tN$ are taken, where $N$ is the memory size, which in this work was to contain 9 concentration reads. So 9 previous concentrations are read to try to predict the next concentration. To forecast particulate matter every 5 minutes, the forecast is made based on data collected in the previous 4 and a half minutes.

\subsection{Architecture and hyperparameters}
\label{arq-hip}

A neural network was then developed using the LSTM model. The model used has only one LSTM hidden layer, as the problem being treated has only one variable, it was decided to use only one layer as it is a simpler problem to deal with. Increasing the number of layers in problems like this may not lead to such considerable gains. The number of nodes/internal neurons ranged between 25 and 100.

Among the activation functions, ReLu, Sigmoid, Tanh and Softmax were used to verify the effectiveness of each one. In addition, the number of epochs was iterated, using 100, 200, 500 and 1000. Only the optmizer Adam was used.

To run the algorithms, we used a computer with an Intel Core i5-4210U 1.70GHz CPU, with integrated graphics, 4GB of RAM and 200GB of storage with Ubuntu 16.04.5 LTS operating system. Tensorflow and Keras libraries \cite{abadi2016tensorflow} were used to implement the neural networks.

\subsection{Results and analysis}

Error metrics such as Mean Absolute Error (MAE), Mean Squared Error (MSE), and Root Mean Squared Error (RMSE) were used to analyze the accuracy of the model. Data were divided into 70\% for training and 30\% for testing. The networks were trained and tested according to the hyperparameters used in Section \ref{arq-hip}.

\begin{figure}[htb]
\centering
\includegraphics[width=0.48\textwidth]{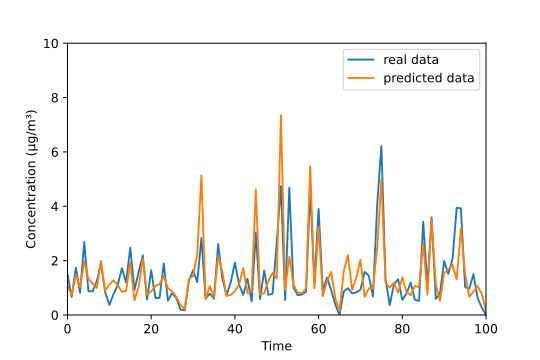}
\caption{Comparison of best model prediction with real data}
\label{modelo-melhor}
\end{figure}

Among these, the one that gave the best result was the one that used the following hyperparameters: 61 internal neurons, Softmax activation function, 100 epochs. The metrics results were: 0.5166 MAE, 1.0234 MSE and 1.0116 RMSE. The MAE evaluates the errors related to the predictions, whereas the MSE is a measure of the estimator's quality, where values close to 0 are better. The RMSE gives us the standard deviation of the residual error, and the smaller the result, the better the performance of the model. Figure \ref{modelo-melhor} presents a comparison of the best model with the real data.

\section{Conclusions}\label{conclusao}



This work presented the proposal of a system (\arquitetura) for monitoring particulate matter on a microscale, through which it is possible to assess the air quality of an environment and to carry out a predictive monitoring of this air quality using neural networks. The system can operate both offline and online.
The \arquitetura~was validated against data collected by meteorological stations. For this purpose, prototype units were installed near IEMA meteorological stations that monitor particulate matter (PM$_{2.5}$). The results showed that low-cost sensor devices (around \$20.00), as is the case in this work, can obtain results similar to those of high-cost meteorological station sensors.

Prototype units were installed in some points in the metropolitan region of Vitória-ES to monitor the air quality, when it was possible to observe that there are differences in the air quality indices (AQI) between neighborhoods, often close. One of the limitations of the prototype is in relation to the environment in which it is being installed, which can increase the amount of particulate material. A future work is to increase the number of units installed in the same neighborhood to observe the dynamics of the AQI and minimize any readings discrepancies in the same region.



The use of low-cost microcontrollers brings limitations, especially in computational power and memory. Using the ESP8266 microcontroller meets the requirements for low cost and Wi-Fi communication. However, it requires a Wi-Fi router for Internet access. In this sense, among the possibilities of evolution of the device is the addition of other communication interfaces, such as LORA and Bluetooth. In the first case, if there is coverage of a LoraWan network, the connection via a Wi-Fi router is avoided. In the second case, the device could transmit the collected data to a cell phone, which, in turn, would transfer them to a server. In both cases, mobile participatory monitoring capacity is added, a desirable property of the system.


As a continuation of the work, it is intended to add to the sensing module the monitoring of other AQI metrics such as temperature, humidity, atmospheric pressure and polluting gases. An ambitious goal that is on the research horizon is to provide the device with the ability to sense liquid particles that contain viruses, such as the Sars-Cov-2, to be used in environments with the possibility of the virus occurring, such as elevators, hospitals and buses. Furthermore, the development of a dashboard for viewing and triggering alerts to authorities is also a future work. Presenting the dynamics of the microscale AQI of the same region can be beneficial both for public policies and for the offer of new services, such as presenting the AQI of the city's regions to those interested in real estate acquisition.



\vspace{-2mm}
\bibliographystyle{sbc}
\bibliography{sbc-template}

\end{document}